\documentclass[epj]{svjour}
\usepackage{amsmath}

\usepackage{bm}
\usepackage{amssymb}
\usepackage{latexsym}
\usepackage{dcolumn}
\newcolumntype{d}{D{.}{.}{-1}}
\ifx\pdftexversion\undefined
  \usepackage[dvips]{graphicx}
\else
  \usepackage[pdftex]{graphicx}
\fi

\begin{document}

%
%
\title{Relativistic correlation correction to the binding energies of the ground configuration
 of Beryllium-like, Neon-like, Magnesium-like and Argon-like ions}

\titlerunning{Relativistic correlation energy\ldots}%
%
\author{J.\ P.\ Santos\inst{1}\and
        G.\ C.\  Rodriges\inst{2,3}\thanks{Deceased}
        J.\ P.\ Marques\inst{2} \and
        F.\ Parente\inst{2} \and
        J.\ P.\ Desclaux\inst{4} \and
        P.\ Indelicato\inst{3}
%
}                     
\offprints{J.\ P.\ Santos}   
\institute{Departamento de F{\'\i}sica, Faculdade de Ci{\^e}ncias e
  Tecnologia, \\
  Universidade Nova de Lisboa, Monte de Caparica, 2825-114 Caparica,
  Portugal, \\
  and Centro de F{\'\i}sica At{\'o}mica da Universidade de Lisboa, \\
  Av. Prof. Gama Pinto 2, 1649-003 Lisboa, Portugal,
  \email{jps@.cii.fc.ul.pt}
   \and Departamento F{\'\i}sica da Universidade de Lisboa \\
   and Centro de F{\'\i}sica At{\'o}mica da Universidade de Lisboa, \\
   Av. Prof. Gama Pinto 2, 1649-003 Lisboa, Portugal,
   \email{parente@cii.fc.ul.pt} 
   \and
 Laboratoire Kastler Brossel,
 \'Ecole Normale Sup\' erieure et Universit\' e P. et M. Curie\\
 Case 74; 4, place Jussieu, 75252 Paris CEDEX 05, France
 \email{paul.indelicato@spectro.jussieu.fr}
 \and
 15 Chemin du Billery
 F-38360 Sassenage,  France 
   \email{jean-paul.desclaux@wanadoo.fr}
  }
\date{Received: \today / Revised version: date }

%
%
\abstract{ Total electronic correlation correction to the binding
  energies of the isoelectronic series of Beryllium, Neon, Magnesium
  and Argon, are calculated in the framework of relativistic
  multiconfiguration Dirac-Fock method. Convergence of the correlation
  energies is studied as the active set of orbitals is increased. The
  Breit interaction is treated fully self-consistently.  The final
  results can be used in the accurately determination of atomic masses
  from highly charged ions data obtained in Penning-trap experiments.
  }

%
%
\maketitle

%
%
\section{Introduction}

The determination of an accurate value for the fine structure constant
$\alpha$ and of accurate mass values has received latelly special
attention due to recent works on highly ionized atoms using Penning
traps~\cite{645,833,832}. The relative uncertainties of such
experimental results can vary from $10^{-7}$ to $10^{-10}$, depending
on the handled ionic species, on the lifetime of the nucleus and on
the experimental apparatus.

When calculating the atomic mass from the experimentally observed ion
mass with this technique, one has to account for the mass $qm_e$ of
the $q$ removed electrons and their atomic binding energy $E_B$.
Therefore, the mass of atom X is given by
\begin{equation}
m_{\text{X}} = m_{\text{X}^{q+}} + qm_e - \frac{E_B}{c^2}
\end{equation}

The influence of the binding energy uncertainties on the mass
determination depends on the specific atom, and increases with the $Z$
value. For example, in the Cs mass determination, an uncertainty of
about 10 eV in the calculated K-, Ar-, and Cl-like Cs ions binding
energies ~\cite{671} originates an uncertainty of the order of
$10^{-11}$ in the mass determination ~\cite{645}.

This means that for the largest uncertainties a simple relativistic
calculated value, in the framework of the Dirac-Fock (DF) approach, is
more than sufficient. However, if the experimental apparatus provides
values with an accuracy that approaches the lower side of the
mentioned interval, one has to perform more sofisticated theoretical
calculations, such as the ones that use the Multi-Configuration
Dirac-Fock (MCDF) model which includes electronic correlation, in
order to achieve a comparable accuracy in the binding energy
determination.

In this article we provide accurate correlation contribution to the
binding energy for the Be-like, Ne-like, Mg-like and Ar-like systems
for atomic numbers up to $Z=95$. We also study self-energy screening effects. The
correlation energies provided here are designed to correct the
Dirac-Fock results of Ref.~\cite{risp2003} for relativistic
correlation effects. In that work, Dirac-Fock energies for all
iso-electronic series with 3 to 105 electrons, and all atomic numbers
between 3 and 118 are provided, using the same electron-electron
interaction operator described in Sec.~\ref{sec:method}.
In Sec.~\ref{sec:method} we give the principle of the calculations,
namely a brief description of the MCDF method used in these
calculations and the enumeration of the radiative corrections
included.
In Sec.~\ref{sec:results} we present the results of calculations and
the conclusions are given in Sec.~\ref{sec:conclusions}.
All numerical results presented here are evaluated with values of the
fundamental constants from the 1998 adjustment \cite{765}.

%
%
\section{Calculations}
\label{sec:method}

To perform theoretical relativistic calculations in atomic systems
with more than one electron, the Brown and Ravenhall
problem~\cite{231}, related to the existence of the $E<-mc^2$
continuum, must be taken in account. To overcome this situation,
Sucher~\cite{229} sugested that a proper form of the electron-electron
interaction with projection operators onto the $E>mc^2$ continuum must
be used, leading to the so called no-pair Hamiltonian,
\begin{equation}
\label{eq:hamilnopai}
        {\cal H}^{\mbox{{\tiny no pair}}}=\sum_{i=1}^{N}{\cal
        H}_{D}(r_{i})+\sum_{i<j}{\cal V}(|\bm{r}_{i}-\bm{r}_{j}|),
\end{equation}
where ${\cal H}_{D}$ is the one electron Dirac operator and 
${\cal V}_{ij} = \Lambda^{++}_{ij} V_{ij} \Lambda^{++}_{ij}$ is an 
operator representing the electron-electron interaction of order
$\alpha$~\cite{47,230}. Here  $\Lambda^{++}_{ij} = \Lambda^{+}_{i} 
\Lambda^{+}_{j}$ is an operator projecting onto the positive energy
Dirac eigenstates to avoid introducing unwanted pair creation effects.
There is no explicit expression for $\Lambda^{++}$, except at the
Pauli approximation~\cite{864}. The elimination of the spurious
contributions from the $E<-mc^2$ continuum in the MCDF
method~\cite{47} is achieved by solving the MCDF radial differential
equations on a finite basis set and keeping in the basis set expansion
only the solutions whose eigenvalues are greater than $-mc^2$ in order
to remove the negative continuum. The basis set used is made of
B-Splines.  The method of Ref.~\cite{47} suffers however from
limitations and inaccuracies due to limitations of the B-Spine basis.
When the number of occupied orbitals is increased, these numerical
errors prevent convergence.  In that case we had to calculate without
projecting. However this problem is not very severe, as the role of
the negative energy continuum becomes less and less important when the
number of electrons increases. In the 4 isoelectronic series studied
here, only the Be-like sequence was sensistive to the presence of the
projection operator even at relatively low $Z$. In the other series,
only the case with $Z=95$ involving the $6h$ shell would have required
it. In the latter case convergence was impossible whether a projection
operator was used or not.

The electron-electron interaction operator $V_{ij}$ is gauge
dependent, and is represented in the Coulomb gauge and in atomic
units, by:
\begin{subequations}
\begin{eqnarray}
\label{eq:eeinter}
         V_{ij}& = & \,\,\,\, \frac{1}{r_{ij}} \label{eq:coulop} \\
         &&-\frac{\bm{\alpha}_{i} \cdot \bm{\alpha}_{j}}{r_{ij}} 
\label{eq:magop} \\ 
         && - \frac{\bm{\alpha}_{i} \cdot
         \bm{\alpha}_{j}}{r_{ij}} 
[\cos\left(\frac{\omega_{ij}r_{ij}}{c}\right)-1]
         \nonumber \\ & & + c^2(\bm{\alpha}_{i} \cdot
         \bm{\nabla}_{i}) (\bm{\alpha}_{j} \cdot
         \bm{\nabla}_{j})
         \frac{\cos\left(\frac{\omega_{ij}r_{ij}}{c}\right)-1}{\omega_{ij}^{2} 
r_{ij}},
         \label{eq:allbreit}
\end{eqnarray}
\end{subequations}
where $r_{ij}=\left|\bm{r}_{i}-\bm{r}_{j}\right|$ is the
inter-electronic distance, $\omega_{ij}$ is the energy of the
exchanged photon between the two electrons, $\bm{\alpha}_{i}$ are
the Dirac matrices and $c=1/\alpha$ is the speed of light.  The term
(\ref{eq:coulop}) represents the Coulomb interaction, the second one
(\ref{eq:magop}) is the Gaunt (magnetic) interaction, and the last two
terms (\ref{eq:allbreit}) stand for the retardation
operator~\cite{845,846}. In the above expression the $\bm{\nabla}$
operators act only on $r_{ij}$ and not on the following wave
functions.
By a series expansion in powers of $\omega_{ij}r_{ij}/c \ll 1$ of the
operators in expressions~(\ref{eq:magop}) and (\ref{eq:allbreit}) one
obtains the Breit interaction, which includes the leading retardation
contribution of order $\alpha^{2}$. The Breit interaction is the sum
of the Gaunt interaction (\ref{eq:magop}) and of the Breit retardation
\begin{equation}
\label{eq:breit}
B^{\mbox{\scriptsize{R}}}_{ij} =
{\frac{\bm{\alpha}_i\cdot\bm{\alpha}_j}{2r_{ij}}} - 
\frac{\left(\bm{\alpha}_i\cdot\bm{r}_{ij}\right)\left(\bm{\alpha}_j
\cdot\bm{r}_{ij}\right)}{2r_{ij}^3}.
\end{equation}

In the present calculation the electron-electron interaction is
described by the sum of the Coulomb and the Breit interaction. The
remaining contributions due to the difference between
Eqs.~(\ref{eq:allbreit}) and (\ref{eq:breit}) were treated only as a
first order perturbation.

%
\subsection{Dirac-Fock method}
\label{subsec:DF}

A first approach in relativistic atomic calculations is obtained
through the relativistic counterpart of the non-relativistic
Hartree-Fock (HF) method, the Dirac-Fock method. The principles
underlying this method are virtually the same as those of the
non-relativistic one.
%
%
In the DF method the electrons are treated in the independent-particle
approximation, and their wave functions are evaluated in the Coulomb
field of the nucleus and the spherically-averaged field from the
electrons. A natural improvement of the method is the generalization
of the electronic field to include other contributions, such as the
Breit interaction.

The major limitation of this method lies in the fact that it makes use
of the spherically-averaged field of the electrons and not of the
local field; i.e., it does not take into account electronic
correlation.

%
\subsection{Multiconfiguration Dirac-Fock method}
\label{subsec:MCDF}

To account for electron
correlation not present at the DF level, one may add to the initial DF configuration,
configurations with the same parity and total angular momentum, involving unoccupied (virtual) orbitals
This is the principle of the Multiconfiguration Dirac-Fock method.

The total energy of an atom, or ion, is the eigenvalue of the
following equation:
\begin{equation}
\label{eq:eq}
        {\cal H}^{\mbox{{\tiny no pair}}}
        \Psi_{\Pi,J,M}(\ldots,\bm{r}_{i},\ldots)=E_{\Pi,J,M}
        \Psi_{\Pi,J,M}(\ldots,\bm{r}_{i},\ldots),
\end{equation}
where $\Pi$ is the parity, $J^2$ is the total angular momentum with
eigenvalue $J$ and its projection on the $z$ axis $J_{z}$, with
eigenvalue $M$. 
The MCDF method is defined by the particular choice of the total wave
function $\Psi_{\Pi,J,M}(...,\bm{ r}_{i},...)$ as a linear
combination of configuration state functions (CSF):
\begin{equation}
\label{eq:totwave}
        \mid\Psi_{\Pi,J,M}\rangle = \sum_{\nu} c_{\nu} \mid\nu \Pi J M
        \rangle.
\end{equation}
The CSF are chosen as eigenfunctions of $\Pi$, $J^2$, and $J_{z}$.  The
label $\nu$ stands for all other numbers (principal quantum number,
coupling, ...)  necessary to define unambiguously the CSF. For a
$N$-electron system, the CSF is a linear combination of Slater
determinants
\begin{equation}
\label{eq:cfs}
        \mid\nu \Pi J M \rangle = \sum_{i} d_{i} \left|
        \begin{array}{ccc} \Phi_{1}^{i}(r_{1}) & \cdots &
        \Phi_{N}^{i}(r_{1}) \\ \vdots & \ddots & \vdots \\
        \Phi_{1}^{i}(r_{N}) & \cdots & \Phi_{N}^{i}(r_{N}) \end{array}
        \right|,
\end{equation}
where the $\Phi$-s are the one-electron wave functions. In the relativistic
case, they are the Dirac four-component spinors:
\begin{equation}
\label{eq:diracspin}
        \Phi_{n \kappa \mu} (\bm{r}) = \frac{1}{r} \left[
        \begin{array}{c} P_{n \kappa}(r) \chi_{\kappa \mu}(\theta ,
        \phi) \\ i Q_{n \kappa}(r) \chi_{-\kappa \mu}(\theta , \phi)
        \end{array} \right]
\end{equation}
where $\chi_{\kappa \mu}(\theta , \phi)$ is a two component Pauli
spherical spinors \cite{78} and $P_{n \kappa}(r)$ and $Q_{n
  \kappa}(r)$ are the large and the small radial components of the
wave function, respectively.  The functions $P_{n \kappa}(r)$, $Q_{n
  \kappa}(r)$ are the solutions of coupled in\-te\-gro-differen\-tial
equations obtained by minimizing Eq.~(\ref{eq:eq}) with respect to
each radial wave function. The coefficients $d_{i}$ are determined
numericaly by requiring that each CSF is an eigenstate of $J^{2}$ and
$J_{z}$, while the coefficients $c_{\nu}$ are determined by
diagonalization of the Hamiltonian matrix (for more details see, e.g.,
Refs.~\cite{31,78,282}).

The  numerical
methods as described in Refs.~\cite{47,282}, enabled the full relaxation of
all orbitals included and the complete self-consistent treatment of
the Breit interaction, i.e., in both the Hamiltonian matrix used for
the determination of the mixing coefficients $c_{\nu}$ in Eq.
(\ref{eq:totwave}) and of the differential equations used to obtain
the radial wave functions. To our knowledge, this is a unique feature
of the MCDF code we used, since others only include the Breit
contribution in the determination of the mixing coefficients (see, e.g.,
\cite{329}).

%
\subsection{Radiative Corrections}
\label{subsec:QED}

The present work is intended to provide correlation energies to
complement the results listed in Ref.~\cite{risp2003}. Radiative
corrections are already included in Ref.~\cite{risp2003}. However, we
give here a discussion of the self-energy screening correction, in
view of a recent work \cite{iam2001}, to compare the uncertainty due
to approximate evaluation of multi-electron QED corrections and those
due to correlation.

The radiative corrections due to the electron-nucleus interaction,
namely the self-energy and the vacuum polarization, which are not
included in the Hamiltonian discussed in the previous sections, can be
obtained using various approximations. Our evaluation, mostly
identical to the one in Ref.~\cite{risp2003} is described as follows.

One-electron self-energy is evaluated using the one-electron results
by Mohr and coworkers~\cite{115,114,116} for several $(n,\ell)$, and
corrected for finite nuclear size~\cite{117}.
Self-energy screening and vacuum polarization are treated with the
approximate method developed by Indelicato and coworkers
~\cite{58,56,53,847}.  These methods yield results in close agreement
with more sophisticated methods based on QED ~\cite{242,263,288}. More
recently a QED calculation of the self-energy screening correction
between electrons of quantum numbers $n\leq 2$, $\ell=0,1$, has been
published \cite{iam2001}, which allows to evaluate the self-energy
screening in the ground state of 2- to 10-electron ions. In the
present work we use these results to evalute the self-energy screening
in Be-like and Ne-like ions.

%
%
\section{Results and Discussion}
\label{sec:results}
 
\subsection{Correlation}
To obtain the uncorrelated energy we start from a Dirac-Fock
calculation, with Breit interaction included self-con\-sis\-ten\-tly.
This correspond to the case in which the expansion~(\ref{eq:totwave})
has only one term in the present work since we study ions with only
closed shells.

The active variational space size is increased by enabling all single
and double excitations from all occupied shells to all virtual
orbitals up to a maximum $n$ and $\ell=n-1$ including the effect of
the electron-electron interaction to all-orders (see ~\cite{671} for
further details). For example, in the Be-like ion case both the $1s$
and $2s$ occupied orbitals are excited up to $2p$, then up to $3d$,
$4f$, $5g$, and $6h$.  We can then compare the difference between
successive correlation energies obtained in this way, to assess the
convergence of the calculation. When calculating correlation
corrections to the \emph{binding} energy it is obviously important to
excite the inner shells, as the correlation contribution to the most
bound electrons provides the largest contribution to the total
correlation energy. However this leads to very large number of
configuration when the number of occupied orbitals is large.  

In the present calculations we used a virtual space spanned by all
singly and doubly-excited configurations. For the single excitations
we excluded the configurations in which the electron was excited to an
orbital of the same $\kappa$ as the initial orbital (Brillouin
orbitals).  In the present case, where there is only one $jj$
configuration in the reference state, those excitations do not change
the total energy, according to the Brillouin theorem (see, e.g.,
\cite{bauche:72,godefroid:1987:brillouin,froese:00}). That would not
be true in cases with open shells in the reference state as it was
recently demonstrated \cite{ild2005}.
The choice of single and double substitutions is due to computation
reasons and is justified by the overwhelming weight of these
contributions.

For all iso-electronic sequences considered here, we included all
configurations with active orbitals up to $6h$, except sometimes for
the neutral case or for $Z=95$, for which convergence problems were
encountered.  The generation of the multiconfiguration expansions was
automatically within the \emph{mdfgme} code.  The latest version can
generate all single and double excitations from all the occupied
levels in a given configuration to a given maximum value of the
principal and angular quantum numbers.  The number of configurations
used to excite all possible pairs of electrons to the higher virtual
orbitals considered is shown in Table~\ref{tab:configuration_number}.
This table shows the rapid increase of the number of configurations
with the number of electrons.

%
%
\begin{table}                                               
\begin{center}                                              
\caption{Number of $jj$ configurations within a given virtual space identified by the correlation orbital with the highest $(n,\ell)$ quantum numbers.}                                             
\label{tab:configuration_number}                                              
\begin{tabular}[c]{lccccc}     
\hline                                              
\hline                                              
        &    $2p$ &    $3d$ &    $4f$ &    $5g$ & $6h$ \\                     
\hline                                 
Be-like &    8    &    38   &    104  &    218  &  392 \\             
Ne-like &         &    84   &    386  &   1007  & 2039 \\             
Mg-like &         &    84   &    486  &   1359  & 2838 \\             
Ar-like &         &    56   &    712  &   2422  & 5505 \\             
\hline                                      
\hline                                      
\end{tabular}                                       
\end{center}                                     
\end{table}

In Table~\ref{tab:corr_be_det} we provide a detailed study of the
contributions to the correlation energy of Be-like ions for $Z$ in the
range $4\leq Z\leq 95$.  We compare several cases. In the first case
the Coulomb correlation energy is evaluated using only the operator
given by Eq. (\ref{eq:coulop}). In the second case, the wavefunctions
are evaluated with the same operator in the SCF process, and used to
calculate the mean-value of the Breit operator (\ref{eq:breit}).
Finally, we include the Breit operator both in the differential
equation used to evaluate the wavefunction (Breit SC) and in the
Hamiltonian matrix. For high-$Z$, relativistic corrections dominate
the correlation energy, which no longer behaves as $A + B/Z+...$, as
is expected in a non-relativistic approximation. The contribution from
the Breit operator represents 34~\% of the Coulomb contribution. It is
thus clear that any calculation claiming sub-eV accuracy must include
the effect of Breit correlation. Obviously higher-order QED effects,
not obtainable by an Hamiltonian-based formalism, can have a similar
order of magnitude.

\begin{table}         
\begin{center}         
\caption{Details of the results for the correlation energy of Be-like
  ions as a function of the operator used in the evaluation of the
  wavefunction and of the size of the active space (see explanations
  in the text). ``all $\to n\ell$'': double excitations from all
  occupied orbitals to all shells up to $n\ell$ are included.}
\label{tab:corr_be_det}         
\scriptsize         
\begin{tabular}[c]{crrrrrr}         
\hline         
\hline         
        &       \multicolumn{6}{c}{Coulomb Correlation, Coulomb SC}             \\
\hline
$Z$     &       $2s^2+2^p2$     &       all $\to 2p$    &       all $\to 3d$    &       all $\to 4f$    &       all $\to 5g$    &           \\
4       &       -1.192  &       -1.192  &       -2.172  &       -2.306  &       -2.392  &               \\
10      &       -3.323  &       -3.328  &       -4.364  &       -4.586  &       -4.688  &               \\
15      &       -4.867  &       -4.876  &       -5.939  &       -6.171  &       -6.274  &               \\
18      &       -5.710  &       -5.720  &       -6.796  &       -7.031  &       -7.136  &               \\
25      &       -7.340  &       -7.353  &       -8.457  &       -8.700  &       -8.807  &               \\
35      &       -8.755  &       -8.774  &       -9.921  &       -10.176 &       -10.286 &               \\
45      &       -9.399  &       -9.427  &       -10.618 &       -10.887 &       -11.000 &               \\
55      &       -9.741  &       -9.778  &       -11.016 &       -11.299 &       -11.417 &               \\
65      &       -10.013 &       -10.057 &       -11.351 &       -11.649 &       -11.775 &               \\
75      &       -10.273 &       -10.321 &       -11.689 &       -12.007 &       -12.142 &               \\
85      &       -10.556 &       -10.607 &       -12.078 &       -12.421 &       -12.568 &               \\
95      &       -11.042 &       -11.094 &       -12.717 &       -13.095 &       -13.257 &               \\
\hline
        &       \multicolumn{6}{c}{Total Correlation, Coulomb SC}       \\
\hline
$Z$     &       $2s^2+2^p2$     &       all $\to 2p$    &       all $\to 3d$    &       all $\to 4f$    &       all $\to 5g$    &          \\
4       &       -1.192  &       -1.192  &       -2.176  &       -2.310  &       -2.396  &               \\
10      &       -3.325  &       -3.330  &       -4.390  &       -4.617  &       -4.722  &               \\
15      &       -4.873  &       -4.882  &       -6.003  &       -6.246  &       -6.357  &               \\
18      &       -5.720  &       -5.731  &       -6.890  &       -7.142  &       -7.257  &               \\
25      &       -7.367  &       -7.382  &       -8.648  &       -8.923  &       -9.048  &               \\
35      &       -8.813  &       -8.835  &       -10.298 &       -10.612 &       -10.753 &               \\
45      &       -9.480  &       -9.513  &       -11.217 &       -11.575 &       -11.734 &               \\
55      &       -9.829  &       -9.874  &       -11.863 &       -12.269 &       -12.445 &               \\
65      &       -10.103 &       -10.159 &       -12.483 &       -12.933 &       -13.147 &               \\
75      &       -10.381 &       -10.446 &       -13.172 &       -13.678 &       -13.926 &               \\
85      &       -10.720 &       -10.794 &       -14.011 &       -14.585 &       -14.878 &               \\
95      &       -11.308 &       -11.392 &       -15.236 &       -15.897 &       -16.245 &               \\
\hline
        &       \multicolumn{6}{c}{Total Correlation, Breit SC}        \\
\hline
$Z$     &       $2s^2+2^p2$     &       all $\to 2p$    &       all $\to 3d$    &       all $\to 4f$    &       all $\to 5g$    &       all $\to 6h$    \\
4       &       -1.192  &       -1.192  &       -2.176  &       -2.310  &       -2.396  &               \\
10      &       -3.325  &       -3.330  &       -4.406  &       -4.616  &       -4.723  &       -4.759  \\
15      &       -4.873  &       -4.882  &       -6.004  &       -6.245  &       -6.360  &       -6.380  \\
18      &       -5.721  &       -5.732  &       -6.887  &       -7.136  &       -7.254  &       -7.303  \\
25      &       -7.367  &       -7.382  &       -8.658  &       -8.936  &       -9.066  &       -9.122  \\
35      &       -8.814  &       -8.836  &       -10.334 &       -10.659 &       -10.811 &       -10.879 \\
45      &       -9.483  &       -9.517  &       -11.308 &       -11.692 &       -11.870 &       -11.951 \\
55      &       -9.836  &       -9.884  &       -12.048 &       -12.502 &       -12.710 &       -12.805 \\
65      &       -10.116 &       -10.179 &       -12.813 &       -13.353 &       -13.594 &       -13.707 \\
75      &       -10.402 &       -10.481 &       -13.712 &       -14.355 &       -14.638 &       -14.771 \\
85      &       -10.750 &       -10.847 &       -14.843 &       -15.618 &       -15.951 &       -16.109 \\
95      &       -11.349 &       -11.473 &       -16.467 &       -17.415 &       -17.812 &               \\\hline         
\hline         
\end{tabular}
\end{center}
\end{table}


In Tables \ref{tab:corr_ne} to \ref{tab:corr_ar} we list the
correlation energy for the Ne-, Mg- and Ar-like sequence with fully
self-consistent Breit interaction, for different sizes of the active
space.  Double excitations from all occupied orbitals to all possible
shells up to $3d$, $4f$, $5g$ and $6h$ are included, except when it
was not possible to reach convergence.

\begin{table*}         
\begin{center}         
\caption{Calculated total correlation energy for the Ne sequence, for different sets of SCF.
  ``all $\to n\ell$'': double excitations from all occupied orbitals
  to all shells up to $n\ell$ are included. Results with Breit self
  consistent included in the calculation. }
\label{tab:corr_ne}         
\scriptsize         
\begin{tabular}[c]{lrrrr}         
\hline         
\hline         
$Z$     &       all $\to 3d$    &       all $\to 4f$    &       all $\to 5g$    &       all $\to 6h$    \\
\hline         
10      &       -5.911  &       -8.306  &       -9.339  &       -9.709  \\
15      &       -5.989  &       -8.712  &       -9.838  &       -10.280 \\
25      &       -6.374  &       -9.310  &       -10.494 &       -10.967 \\
35      &       -6.710  &       -9.850  &       -11.099 &       -11.609 \\
45      &       -7.074  &       -10.482 &       -11.816 &       -12.372 \\
55      &       -7.515  &       -11.269 &       -12.710 &       -13.322 \\
65      &       -8.067  &       -12.260 &       -13.833 &       -14.511 \\
75      &       -8.752  &       -13.375 &       -15.247 &       -16.007 \\
85      &       -9.772  &       -15.119 &       -17.056 &       -17.916 \\
95      &       -11.160 &       -17.231 &       -19.429 &       -20.415 \\
105     &       -13.061 &       -20.129 &       -22.689 &               \\
\hline         
\hline         
\end{tabular}
\end{center}
\end{table*}


\begin{table}         
\begin{center}         
\caption{Details of the results for the correlation energy of Mg-like
  ions as a function of the operator used in the evaluation of the
  wavefunction and of the size of the active space (see explanations
  in the text).  ``all $\to n\ell$'': double excitations from all
  occupied orbitals to all shells up to $n\ell$ are included.}
\label{tab:corr_mg}         
\scriptsize         
\begin{tabular}[c]{crrrr}         
\hline         
\hline         
    &       \multicolumn{4}{c}{Coulomb Correlation, Coulomb SC}        \\
\hline
$Z$ & all $\to 3d$ & all $\to 4f$ & all $\to 5g$ & all $\to 6h$ \\
12 & -3.372 & -7.823 & -9.741 &  \\
20 & -5.211 & -9.724 & -11.809 & -12.640 \\
25 & -5.878 & -10.470 & -12.582 & -13.442 \\
35 & -6.852 & -11.588 & -13.768 & -14.638 \\
45 & -7.477 & -12.349 & -14.597 & -15.477 \\
55 & -7.845 & -12.810 & -15.185 & -16.071 \\
65 & -8.063 & -13.179 & -15.642 & -16.541 \\
75 & -8.220 & -13.596 & -16.081 & -17.001 \\
85 & -8.378 & -14.006 & -16.598 & -17.550 \\
95 & -8.597 & -14.560 & -17.307 & -18.305 \\
\hline
        &       \multicolumn{4}{c}{Total Correlation, Coulomb SC}       \\
\hline
$Z$ & all $\to 3d$ & all $\to 4f$ & all $\to 5g$ & all $\to 6h$ \\
12 & -3.379 & -7.836 & -9.786 &  \\
20 & -5.241 & -9.792 & -11.971 & -12.833 \\
25 & -5.932 & -10.598 & -12.855 & -13.768 \\
35 & -6.977 & -11.891 & -14.351 & -15.338 \\
45 & -7.702 & -12.902 & -15.614 & -16.693 \\
55 & -8.200 & -13.668 & -16.753 & -17.933 \\
65 & -8.580 & -14.424 & -17.879 & -19.172 \\
75 & -8.934 & -15.374 & -19.112 & -20.523 \\
85 & -9.328 & -16.401 & -20.572 & -22.117 \\
95 & -9.833 & -17.718 & -22.422 &  \\
\hline
        &       \multicolumn{4}{c}{Total Correlation, Breit SC}        \\
\hline
$Z$ & all $\to 3d$ & all $\to 4f$ & all $\to 5g$ & all $\to 6h$ \\
12 & -3.379 & -7.864 & -9.734 &  \\
20 & -5.241 & -9.793 & -11.972 & -12.830 \\
25 & -5.932 & -10.599 & -12.857 & -13.762 \\
35 & -6.976 & -11.899 & -14.356 & -15.325 \\
45 & -7.701 & -12.932 & -15.611 & -16.676 \\
55 & -8.198 & -13.808 & -16.799 & -17.939 \\
65 & -8.577 & -14.677 & -18.009 & -19.247 \\
75 & -8.928 & -15.670 & -19.375 & -20.772 \\
85 & -9.319 & -16.939 & -21.062 &  \\
95 & -9.826 & -18.695 & -23.244 &  \\
\hline         
\hline         
\end{tabular}
\end{center}
\end{table}


\begin{table}         
\begin{center}         
\caption{Details of the results for the correlation energy of Ar-like
  ions as a function of the size of the active space (see explanations
  in the text). ``all $\to n\ell$'': double excitations from all
  occupied orbitals to all shells up to $n\ell$ are included. Results
  with Breit self consistent included in the calculation.}
\label{tab:corr_ar}         
\scriptsize         
\begin{tabular}[c]{crrrr}         
\hline         
\hline
$Z$     &       all $\to 3d$    &       all $\to 4f$    &       all $\to 5g$    &       all $\to 6h$    \\
\hline
18      &       -3.258  &       -10.462 &       -13.886 &               \\
20      &       -4.003  &       -11.700 &       -15.203 &       -17.557 \\
25      &       -5.441  &       -13.851 &       -17.755 &       -19.994 \\
35      &       -7.689  &       -16.982 &       -21.292 &       -23.578 \\
45      &       -9.482  &       -19.441 &       -24.093 &       -26.472 \\
55      &       -10.844 &       -21.455 &       -26.486 &       -28.985 \\
65      &       -11.746 &       -23.077 &       -28.564 &       -31.213 \\
75      &       -12.197 &       -24.380 &       -30.426 &       -33.257 \\
85      &       -12.254 &       -25.499 &       -32.229 &       -35.278 \\
95      &       -12.002 &       -26.644 &       -34.207 &               \\
\hline         
\hline         
\end{tabular}
\end{center}
\end{table}


%

\begin{figure}[htb]
\centering

\includegraphics[clip=true,width=5cm,angle=-90]{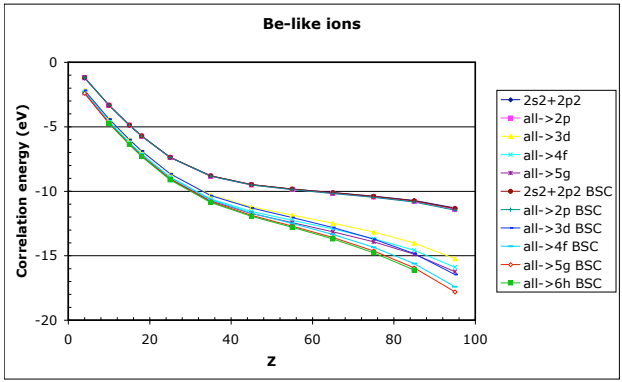}
\caption{Evolution of the
  correlation energy $E_c$ (in eV) for Be-like ions, defined by the diference between
  the total binding energy obtained with the MCDF method and the one
  obtained by the DF method, with the increase of virtual space.}
\label{fig:be-corr}
\end{figure}

\begin{figure}[htb]
\centering
\includegraphics[clip=true,width=5cm,angle=-90]{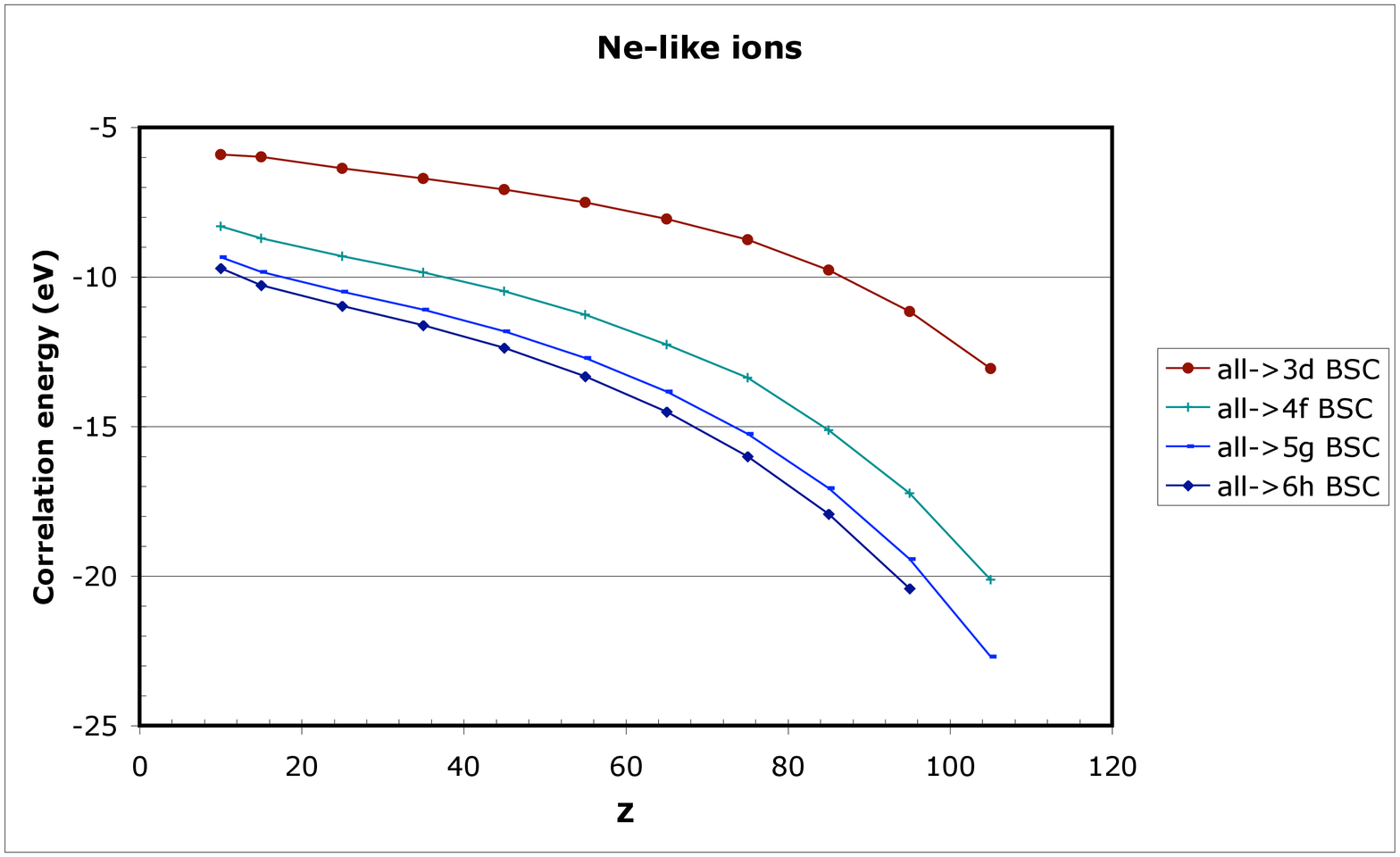}
\caption{Evolution of the
  correlation energy $E_c$ (in eV) for Ne-like ions, defined by the diference between
  the total binding energy obtained with the MCDF method and the one
  obtained by the DF method, with the increase of virtual space.}
\label{fig:ne-corr}
\end{figure}

\begin{figure}[htb]
\centering
\includegraphics[clip=true,width=5cm,angle=-90]{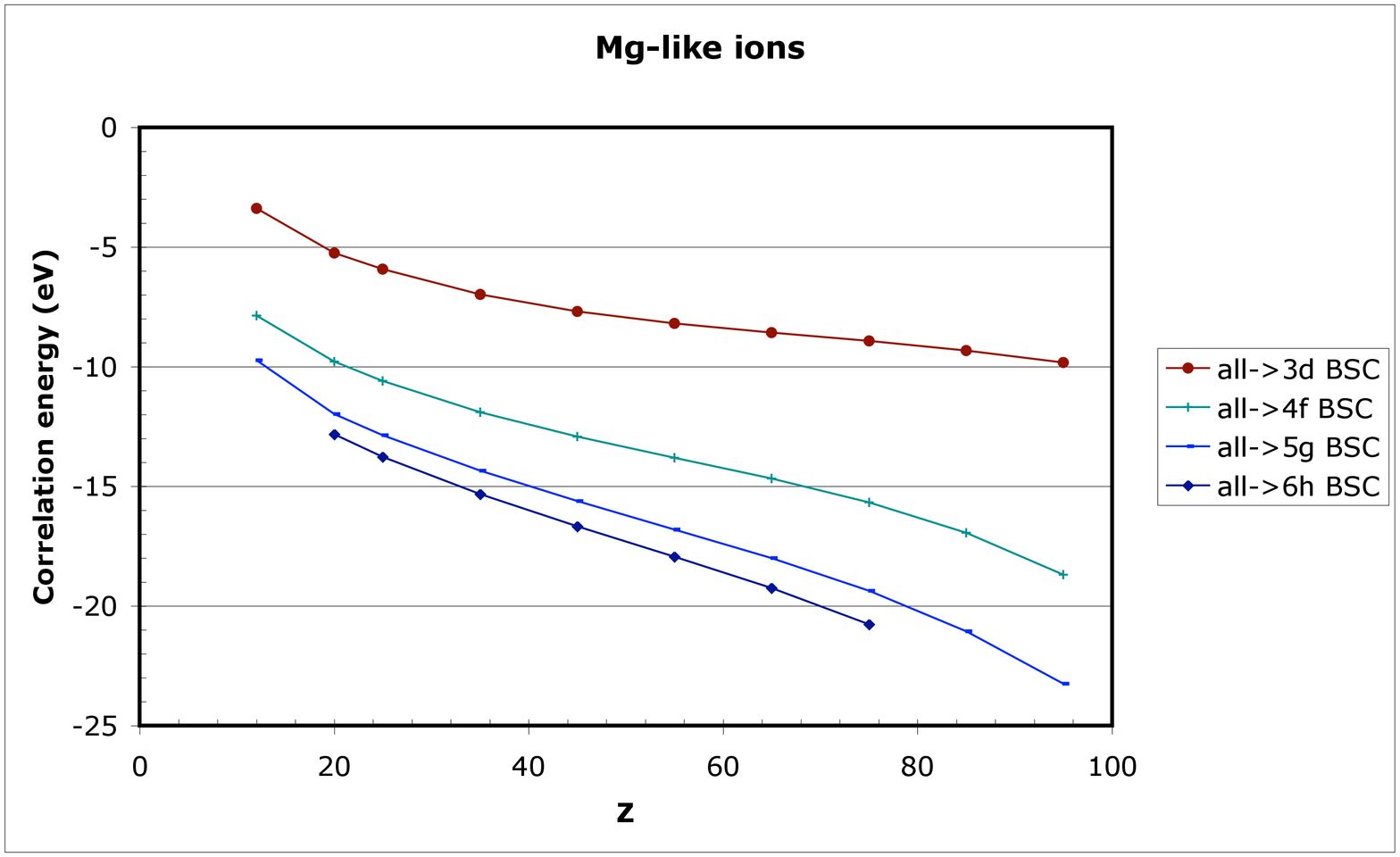}
\caption{Evolution of the
  correlation energy $E_c$ (in eV) for Mg-like ions, defined by the diference between
  the total binding energy obtained with the MCDF method and the one
  obtained by the DF method, with the increase of virtual space.}
\label{fig:mg-corr}
\end{figure}

\begin{figure}[htb]
\centering
\includegraphics[clip=true,width=5cm,angle=-90]{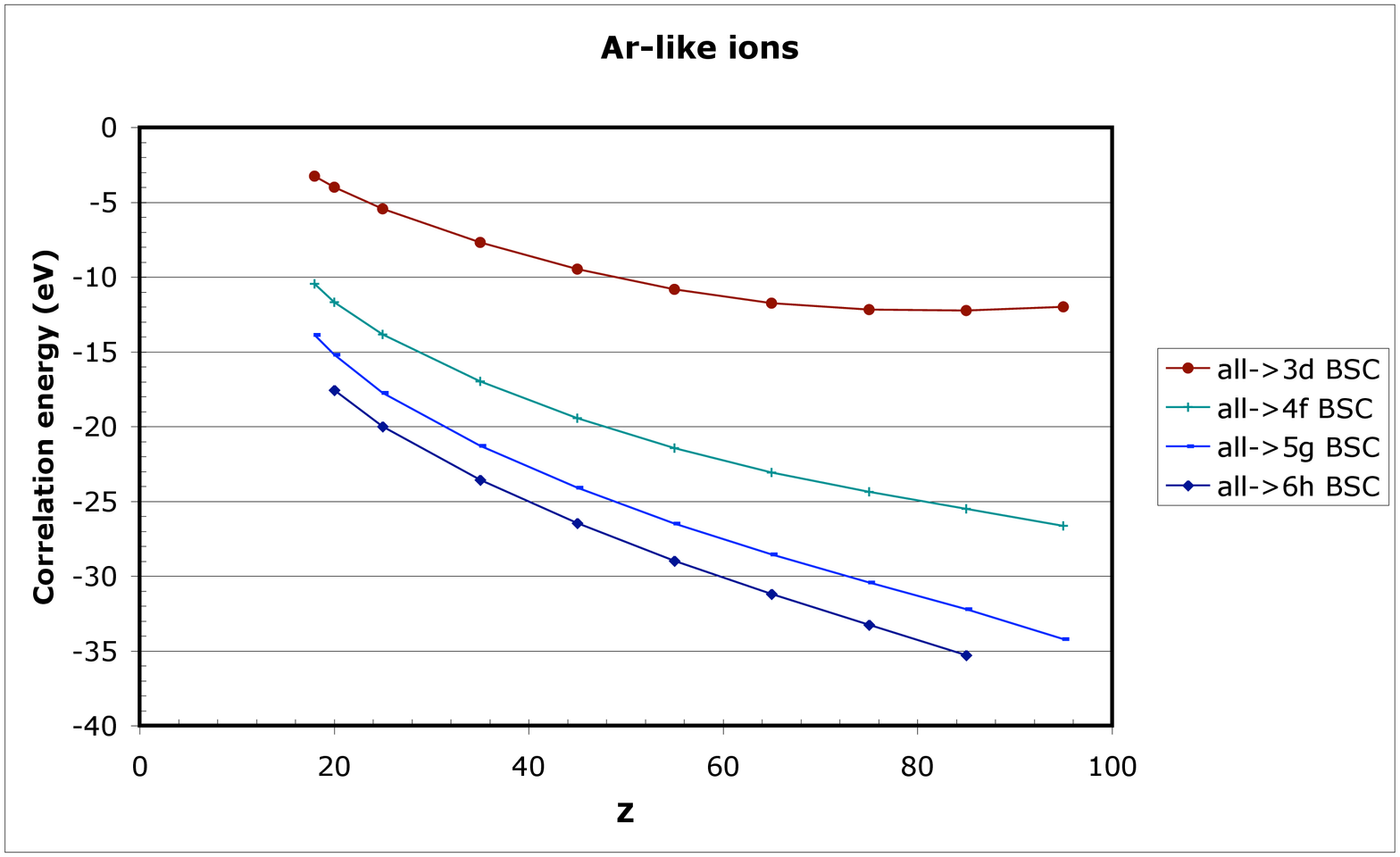}
\caption{Evolution of the
  correlation energy $E_c$ (in eV) for Ar-like ions, defined by the diference between
  the total binding energy obtained with the MCDF method and the one
  obtained by the DF method, with the increase of virtual space.}
\label{fig:ar-corr}
\end{figure}

In Fig.~\ref{fig:be-corr} to \ref{fig:ar-corr} we present the
evolution of the correlation energy $E_c$ (in eV), defined by the
diference between the total binding energy obtained with the MCDF
method and the one obtained by the DF method, with the increase of the
virtual space for each isoelectronic series studied. We notice, as
expected, a decrease of the energy with the increase of the atomic
number and the increase of the number of virtual orbitals.


An inspection of Fig.~\ref{fig:be-corr} to \ref{fig:ar-corr} and of
Tables \ref{tab:corr_ne} to \ref{tab:corr_ar} gives a clear indication
of the importance of including a specific shell in the calculation for
the value of the correlation, i.e., if a new curve, corresponding to
the inclusion of a specific shell, is close to the previous curve,
obtained through the inclusion of shells of lower principal quantum
number, it means that we have included the major part of the
correlation in the energy calculation. We can also see the effect of
including or not the Breit interaction in the SCF process.  Our
calculation is accurate within a few 0.01~eV for low-$Z$ Be-like ions
up to 0.15~eV at high-$Z$. For Ne-like ions, we find respectively
0.4~eV and 1~eV, for Mg-like ions we find 0.9 and 1.4~eV, and for
Ar-like ions these numbers are 2.3 and 3~eV. It is thus clear that the
maximum value of $n$ and $\ell$ one should go to reach uniform
accuracy increases with the number of electrons. However the
uncertainty due to this limitation of our calculation is probably
negligible compare to neglected QED corrections like the contribution
from negative energy continuum, box diagram and two-loop QED
corrections.

In order to provide values for arbitrary atomic numbers within each
isoelectronic series we have fitted polynomials to the best
correlation curves. The equations for these fits are given in Table
\ref{tab:fit}.

\begin{table*}[htbp]
  \centering
  \caption{Fit to the ground state total correlation energy $ \Delta
    E$ of the Be, Ne, Mg and Ar isoelectronic sequences, with correlation orbitals up to $6h$}
\label{tab:fit}
 \begin{tabular}{cl}
\hline
\hline
Series & \multicolumn{1}{c}{Fit} \\
\hline
 Be &$ \Delta E = 1.421329\times 10^{-7}Z^4 - 7.019909\times 10^{-5}Z^3 + 9.159169\times 10^{-3}Z^2 - 5.474933\times 10^{-1}Z - 7.191674\times 10^{-2}$\\
Ne & $ \Delta E = 5.523943\times 10^{-8}Z^4 - 2.760868\times 10^{-5}Z^3 + 2.214132\times 10^{-3}Z^2 - 1.324244\times 10^{-1}Z - 8.627745
$\\
Mg & $\Delta E = -2.156149\times 10^{-8}Z^4 - 1.529410\times 10^{-5}Z^3 + 2.928077\times 10^{-3}Z^2 - 2.903759\times 10^{-1}Z - 8.078404$\\
Ar &$\Delta E = 5.696195\times 10^{-7}Z^4 - 1.529548\times 10^{-4}Z^3 + 1.589991\times 10^{-2}Z^2 - 9.710181\times 10^{-1}Z - 3.406304$
 \\
\hline
\hline
  \end{tabular}
  
\end{table*}


We present in Table~\ref{tab:contributions_be} the different terms
contributing to the total atomic binding energy of Be-like ions with
$Z=4$, 45 and 85, to illustrate their relative importance.

%
%
\begin{table*}         
\begin{center}         
\caption{Contributions to the atomic binding energy for
    for ions of different $Z$ in the Beryllium isoelectronic serie (in
    eV).}
\label{tab:contributions_be}         
\scriptsize         
\begin{tabular}[c]{lrrr}         
\hline         
\hline         
 & Z=4 & Z=45 & Z=85  \\
\hline         
 &  &  &   \\
Coulomb & -398.91260 & -68961.32493 & -272463.59996  \\
Magnetic & 0.01430 & 39.84888 & 310.21457  \\
Retardation (order $\omega^2$) & 0.00105 & -0.58860 & -6.10695  \\
Higher-order retardation ($>\omega^2$) & 0.00000 & 0.00000 & 0.00000  \\
Hydrogenlike self-energy & 0.01310 & 62.62419 & 610.43890  \\
Self-energy screening & -0.00291 & -1.76962 & -13.44919 \\
Vacuum polarization (Uheling) $\alpha(Z\alpha)$ & -0.00039 & -7.46054 & -139.37727  \\
Electronic correction to Uheling & 0.00004 & 0.03290 & 0.33323  \\
Vacuum polarization $\alpha(Z\alpha)^3$ & 0.00000 & 0.12368 & 6.14067 \\
Vac. Pol. (K{\"a}ll{\`e}n \& Sabry) $\alpha^2(Z\alpha)$ & 0.00000 & -0.06042 & -1.07200  \\
Recoil & 0.00000 & -0.00805 & -0.06221  \\
Correlation & -2.39600 & -11.95100 & -16.10900 \\
Total Energy & -401.28341 & -68880.53351 & -271712.6492  \\
\hline         
\hline         
\end{tabular}
\end{center}
\end{table*}


\subsection{Self-energy screening}
In Table \ref{tab:qed} we compare the self-energy screening correction
evaluated by the use of Ref. \cite{iam2001} and by the Welton method.
Direct evaluation of the screened self-energy diagram using
Ref.~\cite{iam2001}, includes relaxation only at the one-photon
exchange level.  The Welton method include relaxation at the
Dirac-Fock or MCDF level. In the case of Be-like ions we also
performed a calculation including intra-shell correlation to have an
estimate of the effect of correlation on the self-energy screening.
The change due to the method is much larger than the effect of even
strong intra-shell correlation. The difference between the two
evaluations of the self-energy screening can reach $\approx 2$~eV at
Z=95.  

\begin{table*}         
\begin{center}         
\caption{Comparison of the screened self-energy contribution in
  Be-like and Ne-like ions obtained by different methods.}
\label{tab:qed}         
\scriptsize         
\begin{tabular}[c]{crrrrrr}         
\hline         
\hline         
        &       \multicolumn{4}{c}{Be-like}     &       \multicolumn{2}{c}{Ne-like}                     \\
        &       \multicolumn{2}{c}{Ref. \cite{iam2001}} &\multicolumn{2}{c}{Welton model}&     \multicolumn{1}{c}{Ref. \cite{iam2001}}&\multicolumn{1}{c}{Welton model}\\
Z       &       $2s^2$  &       $2s^2+2p^2$     &       $2s^2$  &       $2s^2+2p^2$     &               &               \\
\hline         
4       &       -0.004  &       -0.004  &       -0.003  &       -0.003  &               &               \\
10      &       -0.047  &       -0.046  &       -0.036  &       -0.035  &       -0.081  &       -0.050  \\
15      &       -0.132  &       -0.129  &       -0.104  &       -0.101  &       -0.229  &       -0.155  \\
25      &       -0.466  &       -0.458  &       -0.384  &       -0.375  &       -0.835  &       -0.614  \\
35      &       -1.066  &       -1.053  &       -0.917  &       -0.903  &       -1.973  &       -1.519  \\
45      &       -1.995  &       -1.976  &       -1.801  &       -1.783  &       -3.825  &       -3.060  \\
55      &       -3.349  &       -3.323  &       -3.190  &       -3.165  &       -6.659  &       -5.530  \\
65      &       -5.282  &       -5.248  &       -5.317  &       -5.279  &       -10.888 &       -9.388  \\
75      &       -8.054  &       -8.012  &       -8.562  &       -8.499  &       -17.178 &       -15.373 \\
85      &       -12.130 &       -12.080 &       -13.546 &       -13.439 &       -26.659 &       -24.737 \\
95      &       -19.176 &       -19.109 &       -21.347 &       -21.162 &       -41.114 &       -39.721 \\
\hline         
\end{tabular}
\end{center}
\end{table*}


%
%
%

\section{Conclusions}
\label{sec:conclusions}

We have presented relativistic calculations of the correlation
contribution to the total binding energies for ions of the Beryllium,
Neon, Magnesium and Argon isoelectronic series.  We have shown that
accurate results can be achieved if excitations to all shells up to
the $n=6$ shell are included.We have also compared two different
methods for the evaluation of the self-energy screening.  Combined
with the results of Ref.~\cite{risp2003} our results will provide
binding energies with enough accuracy for all ion trap mass
measurements to come, involving ions with the isolelectronic sequences
considered here.

%
%
%
\section*{acknowledgments}
This research was partially supported by the FCT project
POCTI/FAT/50356/2002 financed by the European Community Fund FEDER,
and by the TMR Network Eurotraps Contract Number ERBFMRXCT970144.
Laboratoire Kastler Brossel is Unit{\'e} Mixte de Recherche du CNRS
n$^{\circ}$ C8552.
%
%

%
%
\bibliographystyle{epjdsty}
\bibliography{jps}

\end{document}